\documentclass[a4paper]{mem}
\usepackage{txfonts}                                                                      
\usepackage{natbib}
\usepackage{graphicx}
\usepackage[a4paper]{hyperref}
\idline{xx}{1}

\begin{document}

%
\def\etal{et  al.\ }
\def\araa{{Ann.\ Rev.\ Astron.\ Ap.}}
\def\aplet{{Ap.\ Letters}}
\def\aj{{Astron.\ J.}}
\def\apj{ApJ}
\def\apjl{{ApJ\ (Lett.)}}
\def\apjs{{ApJ\ Suppl.}}
\def\aas{{Astron.\ Astrophys.\ Suppl.}}
\def\aa{{A\&A}}
\def\aap{{A\&A}}
\def\mnras{{MNRAS}}
\def\nat{{Nature}}
\def\pasa{{Proc.\ Astr.\ Soc.\ Aust.}}
\def\pasp{{P.\ A.\ S.\ P.}}
\def\pasj{{PASJ}}
\def\pre{{Preprint}}
\def\sovlet{{Sov. Astron. Lett.}}
\def\adspr{{Adv. Space. Res.}}
\def\expas{{Experimental Astron.}}
\def\ssr{{Space Sci. Rev.}}
\def\apss{{Astrophys. and Space Sci.}}
\def\inpress{in press.}
\def\souspresse{sous presse.}
\def\inprep{in preparation.}
\def\enprep{en pr\'eparation.}
\def\submit{submitted.}
\def\soumis{soumis.}
\def\aph{{Astro-ph}}
\def\astroph{{Astro-ph}}

%
\def\keV {\rm keV}
\def\msol{{M$_{\odot}$}}

%
\def\rc{{r_{\rm c}}}
\def \rci {r_{\rm ci}}
\def \rcut {R_{\rm cut}}
\def\ne{{n_{\rm e}}}
\def\nh{{N_{\rm H}}}
\def\kT{{\rm k}T}
\def\rs {r_{\rm s}}
\def \fgas  {f_{\rm gas}}
\def \Mgas  {M_{\rm gas}}
\def \Mv  {M_{\rm 200}}
\def \rv  {R_{200}}
\def \rhoDM {\rho_{\rm DM}}
\def \rhog {\rho_{\rm gas}}
\def \rhoc {\rho_{\rm c}}
\def \rhocz {\rho_{\rm c}(z)}
\def \ne {n_{\rm e}}
\def \Lx {L_{\rm X}}

\def \LxT {\hbox{$\Lx$--$T$} }
\def \MgT {\hbox{$\Mgas$--$T$} } 
\def \MT {\hbox{$M$--$T$} } 
\def \RvT {\hbox{$\rv$--$T$} }  
\def \EMT {\hbox{$EM$--$T$} } 
\def \ST {\hbox{$S$--$T$} } 

\def\hzero{{H_{0}}}
\def\qzero{{q_{0}}}

\def \xmm {\hbox{\it XMM-Newton}}
\def \asca {\hbox {\it ASCA}}
\def \rosat  {\hbox {\it ROSAT}}
\def \chandra  {\hbox {\it Chandra}}
\def \sax {\hbox {\it Beppo-SAX}}
\def\betamod{$\beta$-model}
\def \etal {et al.\ }

%
%
   \title{Structural and scaling properties of galaxy clusters: probing the physics of structure formation}
   \author{M. Arnaud, G. W. Pratt\thanks{Present address: MPE Garching, Giessenbachstra{\ss}e, 85748 Garching, Germany} \and  E. Pointecouteau} 
   \offprints{M. Arnaud}
\mail{marnaud@discovery.saclay.cea.fr}
   \institute{CEA/DSM/DAPNIA/Service d'Astrophysique, CE Saclay,
L'Orme des Merisiers, 91191 Gif sur Yvette, France}
 \abstract{ 
   We present \xmm\ studies of the total mass, gas density, temperature and entropy profiles in nearby hot and cool clusters, together with follow-up observations of distant clusters from the SHARC Survey.  The observed structural and scaling properties are compared with the predictions of the self-similar model of cluster formation. These data indicate that clusters do form a self-similar population down to low mass and up to high redshift, and give support to the standard picture of structure formation for the dark matter component.  However, deviations from the standard scaling laws confirm that the specific physics of the gas component is still insufficiently understood.
     }
   \authorrunning{Arnaud, Pratt \& Pointecouteau }
   \titlerunning{Structural and scaling properties of galaxy clusters}
   \maketitle
%

\section{Introduction}
\subsection{The self-similar model of cluster formation}

In the standard hierarchical formation scenario, clusters of galaxies are forming in the
recent cosmological epoch ($z\sim 2$ to the present time). In theory, X--ray cluster formation and evolution is simple and driven by the collisionless gravitational collapse of the main dark matter (DM) component, whereas the hot gaseous intra-Cluster Medium (ICM) `follows' the gravitational potential of the DM.  Analytical models of the DM collapse (\citealt{bert85};  \citealt{cav99}), supported by numerical simulations (\citealt{eg02} for a review), then predict that galaxy clusters constitute a self-similar population.  

In such a model, purely based on gravitation, the internal shape of clusters is universal, independent of mass and redshift. Furthermore, the virialised part of a cluster, present at a given redshift $z$, corresponds to a fixed density contrast, $\delta_{\rm c}\sim200$, as compared to the critical density of the Universe at that redshift, $\rhocz$: 
\begin{equation}         
\frac{\Mv}{ \rv^3} = \frac{4\pi}{3} \delta_{\rm c}\ \rhocz 
\label{eq:rho} 
\end{equation} 
where $\Mv$ and $\rv$ are the virial mass and radius.  $\rhocz= h^{2}(z) 3 H_0^2/( 8 \pi G)$, where $h(z)$ is the Hubble constant normalized to its local value.  $h(z)$ depends on the cosmological density parameter $\Omega_{\rm m}$ and the cosmological constant $\Lambda$: $h^{2}(z)=\Omega_{\rm m}(1+z)^{3} +\Lambda$. 
 
The ICM is assumed to be an ideal gas evolving in the gravitational potential of the dark matter. The gas mass fraction $\fgas$ is thus constant:
\begin{equation}        
\fgas =  \frac{\Mgas}{\Mv} = cst
\label{eq:fgas} 
\end{equation} 
and self similarity applies to both the DM component and the ICM.  The virial theorem then gives: 
\begin{equation} 
\frac{G \mu {\rm m_{p}}\ \Mv }{2\ \rv}  = \beta_{T}\ \kT
\label{eq:viriel} 
\end{equation} 
where $T$ is the mean temperature and $\beta_{\rm T}$ is a normalisation constant depending on cluster (universal) internal structure.

\begin{figure}[t]
    \hspace*{-1em}\includegraphics[width=7cm]{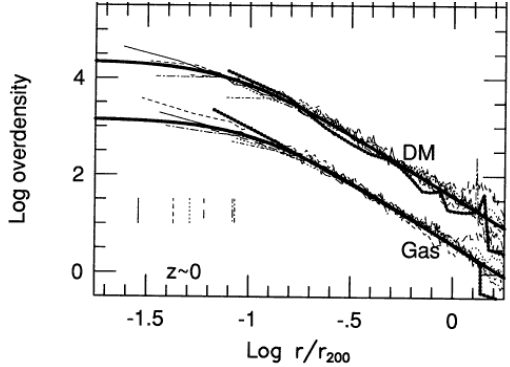}\vspace*{-12em}
    \caption{\footnotesize Universal profiles obtained from numerical simulations of the formation and evolution of galaxy clusters (from \citealt{nfw95}).  The radius is scaled to the virial radius, corresponding to an overdensity of 200 as compared to the mean density of the Universe in which the cluster is embedded.   The Dark matter and gas densities are normalized to the mean density.   }
 \label{frhoprof}%
    \end{figure}
 
 From the basic equations Eq.~\ref{eq:rho}, Eq.~\ref{eq:fgas} and Eq.~\ref{eq:viriel}, it is easy to see that the cluster population is a two parameter population. Simple scaling laws relate each physical property, $Q$, to the cluster X-ray temperature and redshift, in the form $Q\propto A(z)T^{\alpha}$.  The evolution factor,  $A(z)$, is due to the evolution of the mean DM (and thus gas) density, which varies as the critical density of the Universe:
\begin{equation}
  \overline{\rhog} \propto \overline{\rhoDM} \propto \rhocz \propto h^{2}(z)
  \label{eq:rho2}
  \end{equation}
For instance, the total and gas  mass scales as $\Mgas \propto \Mv \propto h^{-1}(z)  T^{3/2}$, the virial radius as $\rv  \propto   h^{-1}(z) T^{1/2}$ and the X-ray luminosity, assuming bremsstrahlung emission, as $\Lx \propto h(z) T^{2}$. 

\subsection{Insights from X-ray observations}
 
How the statistical properties of the observed cluster population compare with these theoretical
predictions is a key issue in modern cosmology.  Such comparisons provide unique information on the physics that governs the formation and evolution of large scale structure, both for the DM and the baryonic components.  

We can first examine relations between global properties such as luminosity, gas mass, total mass, size and temperature. This is a powerful method, and has been much used with data from previous satellites (\citealt{arn02r} for a review and references below).  Although strong correlations are observed between various quantities, it was rapidly recognized that clusters deviate from the classic self-similar model.  The most remarkable deviation is the local \LxT relation, which is steeper than expected (e.g. \citealt{ae99};  \citealt{mark98a}). This was the first  indication that the gas physics should be looked at more closely and that non-gravitational processes could play a role (e.g. \citealt{eh91}).

An unambiguous picture can be obtained only by studying the internal structure of clusters. For instance a steepening of the \LxT relation could be due to a systematic increase of  $\fgas$ with T (a simple modification of the scaling laws) and/or a variation of cluster shape with T (a break of self-similarity).  By looking at radially averaged profiles of interesting quantities (temperature, density, entropy, integrated mass, etc)  two considerations can be addressed at the same time. The first is to see whether the profiles agree in {\it shape\/}; the second is to investigate the {\it scaling\/} of the profiles.  The existence of a universal underlying dark matter profile in the simulations (\citealt{nfw97}; hereafter NFW) spurred searches to detect the observable signatures of such universality in the gas density and temperature profiles. This was indeed found to be the case when \rosat\ and \asca\ data of {\it hot} clusters were examined (e.g. \citealt{mark98b}; \citealt{neum99}).  \rosat/\asca\  observations also gave the first indication that the self-similarity does hold up to high redshift \citep{aan02}.

The present generation of X-ray satellites represents a giant step forward in terms of resolution and sensitivity.  In marked contrast to studies with the previous generation of satellites, we are subject to far less uncertainty on the temperature profiles (and thus on the total mass and entropy profiles) due to PSF effects. Furthermore, the sensitivity of \xmm\ now allows us to probe the scaling and structural properties further away from the cluster centre, down to lower mass clusters  and at higher precision at high  redshifts. 



 \section{Dark matter in nearby clusters}

\subsection{A universal dark matter profile}
The observed shape of the DM distribution is an important test of our understanding of the DM collapse.  We derived dark matter profiles  from the \xmm\ observations of three nearby clusters of galaxies. Two of these are hot and massive: A1413 ($z=0.143$; mean temperature $\kT=6.5$~keV; \citealt{pratt02}) and A478 ($z=0.0881$;  $\kT=6.7$~keV; \citealt{pointeco03}). The third cluster is the cool system A1983 ($z=0.044$),  with $\kT=2.3~\keV$ \citep{pratt03}. The total mass profiles were obtained from the temperature and gas density profiles under the hypothesis of hydrostatic equilibrium and spherical symmetry. Data processing details (background subtraction, vignetting and PSF correction, deprojection) can be found in the papers.  The total mass profiles are well fitted using a dark matter model as described by NFW. In the case of A478 the radial coverage was wide enough (especially in the centre) to distinguish between the NFW and the more centrally peaked profile found in the simulations by \citet{moor99}. Such a profile was found to be inconsistent with our data, as well as profiles without a cusp, like the King profile. 

\begin{figure}[t]
   \hspace*{-1em}\includegraphics[width=7cm]{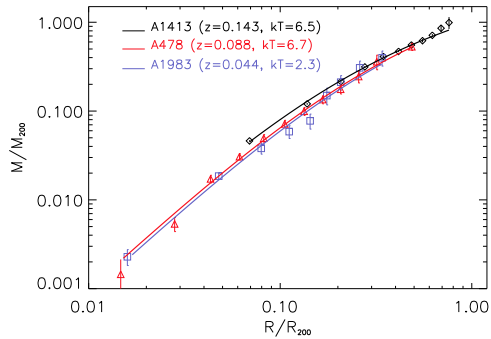}\vspace*{-1em}
    \caption{\footnotesize Integrated total mass of A1413 (black diamonds; \citealt{pratt02}), A478 (red triangles; \citealt{pointeco03}) and A1983 (blue squares; \citealt{pratt03}) scaled  to their virial mass versus the scaled radius (\xmm\ data). The best fit NFW model profile is overplotted in the same color code for each cluster.    }
              \label{fmprof}%
    \end{figure}

The three profiles were scaled according to their virial mass, $M_{200}$,
corresponding to a density contrast  of 200,  
and their virial radius ($\rv$), as shown in Fig.~\ref{fmprof}. The three scaled profiles are in good agreement. From these three \xmm\ observations, we are able to describe the shape of a universal dark matter profile in nearby clusters in the radial range $\sim0.01 \rv$ to $0.7 \rv$. Our result suggests that this profile shape is independent of the cluster temperature (or mass) as predicted in the hierarchical model of structure formation.

\begin{figure}[t]
     \hspace*{-1em}\includegraphics[width=7cm]{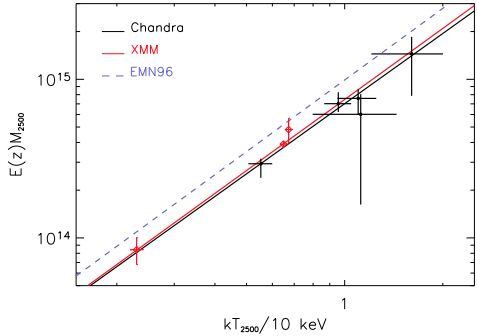}\vspace*{-1em}
   \caption{\footnotesize  $M-T$ relation at $R_{2500}$ as seen by \xmm\ (red data points and solid line) and \chandra\ (black data points and solid line, \citealt{allen02}). The blue dotted line is the prediction from numerical simulations \citep{emn96} 
   }
              \label{fmt}%
    \end{figure}

\subsection{The $M-T$ relation}
\label{sec:mtrel}
A further test of the scaling properties of nearby clusters is provided by the \MT relation.  No definitive answer could be derived from  \asca/\rosat\  observations, various studies providing different scalings in both normalisation and slope.  For instance, it was unclear whether the mass scales as $T^{3/2}$ as expected \citep{horn99}, or if this was true only in the high mass regime ($\kT>3-4~\keV$, see also \citealt{etto02}),  with a steepening at lower mass (\citealt{neva00}, \citealt{xu01}, \citealt{fino01}), or even if the slope was higher all over the mass range \citep{sand03}. This is likely to reflect both the large uncertainties in the temperature profiles (see below), and various extrapolation biases, especially for low mass systems.  

 We computed the total mass of our three clusters at a radius corresponding to a density contrast of 2500 ($M_{2500}$). A power law fit was performed in the $M-T$ plane, leading us to an observational $M-T$ relation as seen by \xmm: $M_{2500}\propto T^{1.49\pm0.2}$. This relation can be compared with the relation derived at the same density contrast by \citet{allen02} from \chandra\ observations of six hot and massive clusters: $M_{2500}\propto T^{1.52\pm0.36}$. Both results are in excellent agreement (Fig.~\ref{fmt}) and are consistent with the standard scaling. The precision on the slope obtained with the \xmm\ data from a sample of only three clusters is remarkable. This is both due to the precision of the individual mass estimates and the unique leverage provided by \xmm\  at  the low mass end.  

However, comparison with the prediction from numerical simulations  \cite{emn96} exhibits a significant discrepancy in terms of normalisation. This discrepancy appears to be present at all radii \citep{pratt02}. Knowing that the normalisation of the $M-T$ relation depends on both the gas density and dark matter profiles, this result strongly suggests that, while the dark matter collapse seems to be correctly understood, we are still unable to model correctly the gas profile in terms of shape and distribution.  In the next section we will discuss the gas properties.

\section{Some insights into cluster gas physics}
\subsection{Emission measure profiles}
It is first useful to consider the emission measure along the line of sight $EM(r) = \int_{r}^{\rv} \ne^{2}~dl$, which is easily derived from the X-ray surface brightness profile and is directly linked to the gas content and density distribution. The scaled $EM$ profiles of {\it hot} clusters were considered by Neumann \& Arnaud (1999; 2001; see also \citealt{vikh99}) and were found to be similar in shape outside the cooling core region. Recall that in the standard self-similar framework, the mean gas  density does not depend on the temperature (see Eq.~\ref{eq:rho2}). The emission measure thus scales as $\rv$, i.e $EM \propto T^{0.5}$. These authors found that the scatter was considerably reduced if a much steeper $EM$---$T$ relation, $EM \propto T^{1.38}$, was instead used. This explains the steepening of the \LxT relation and translates into $M_{\rm gas} \propto T^{1.94}$, consistent with the observed steepening of the $M_{\rm gas}$---$T$ relation (\citealt{mohr99}; see also \citealt{vikh99}; \citealt{etto02}; \citealt{cast03}) as compared to the standard $\Mgas \propto T^{1.5}$ scaling.
 \begin{figure*}
   \centering 
   \hspace*{-1em}\includegraphics[width=12cm]{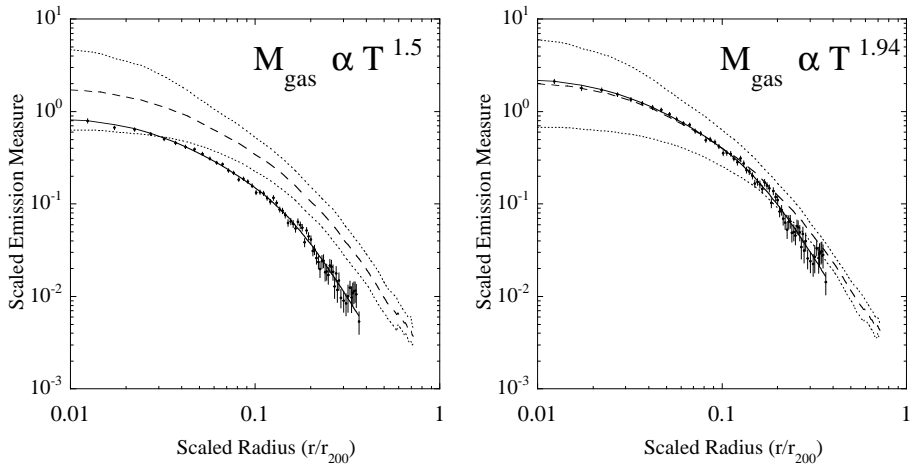}
   \vspace*{-0.5em}
    \caption{\footnotesize The \xmm\ scaled emission measure (EM) profile of A1983 (data points) compared to the mean scaled EM profile of the ensemble of hot clusters ($\kT > 3.5$ keV) observed by \rosat\ and studied  by \citet{neum01}.  The latter is indicated by the dashed line; dotted lines show the dispersion.  On the left, the EM profiles are scaled using the self-similar  scaling $EM \propto T^{0.5}$ (corresponding to $\Mgas \propto T^{1.5}$).  On the right, it is scaled using the empirical $EM \propto T^{1.38}$ ($\Mgas \propto T^{1.94}$) scaling of \citet{neum01}. }
   \label{fig:emprof}
    \end{figure*}

In Fig.~\ref{fig:emprof} we show the $EM$ profile of A1983 ($\kT = 2.13$ keV; \citealt{pratt03}) compared to the mean profile for hotter clusters ($\kT > 3.5$ keV; the envelope indicates typical dispersion about the mean). In the left-hand panel, the profiles are scaled using the standard self-similar scaling, $EM \propto T^{0.5}$. In the right hand panel, the empirically-derived scaling of $EM \propto T^{1.38}$ is used. The use of the non-standard scaling not only reduces the scatter in the mean profile for hotter clusters, it also brings the $EM$ profile of A1983 into the envelope of values for hotter clusters.

This result is interesting because it implies a universal gas density profile shape down to quite low temperature --- the $EM$ profile of A1983 is not flatter than that of hotter clusters. The result also shows that the observed $EM$---$T$ relation is indeed steeper than expected, consistent with the observed steepening of the $M_{\rm gas}$---$T$ relation. 

\subsection{Temperature profiles}
The exact form of cluster temperature profiles is a topic which has generated a huge amount of debate in the literature. Depending on the study, cluster profiles are steeply declining 
\citep{mark98b} [\asca], have a isothermal core beyond which there is a steep decline \citep{dm02}  [\sax]), or are flat to the limits of detection (\citealt{whit00} [\asca]; \citealt{ib00}  [\sax]). The differing conclusions reflect the complexity properly of treating the energy-dependent PSF of these satellites. The PSF of \xmm\ is far less of a problem, being far smaller, and much less energy dependent. 

\begin{figure}[t]
   \hspace*{-1em}
   \includegraphics[width=7cm]{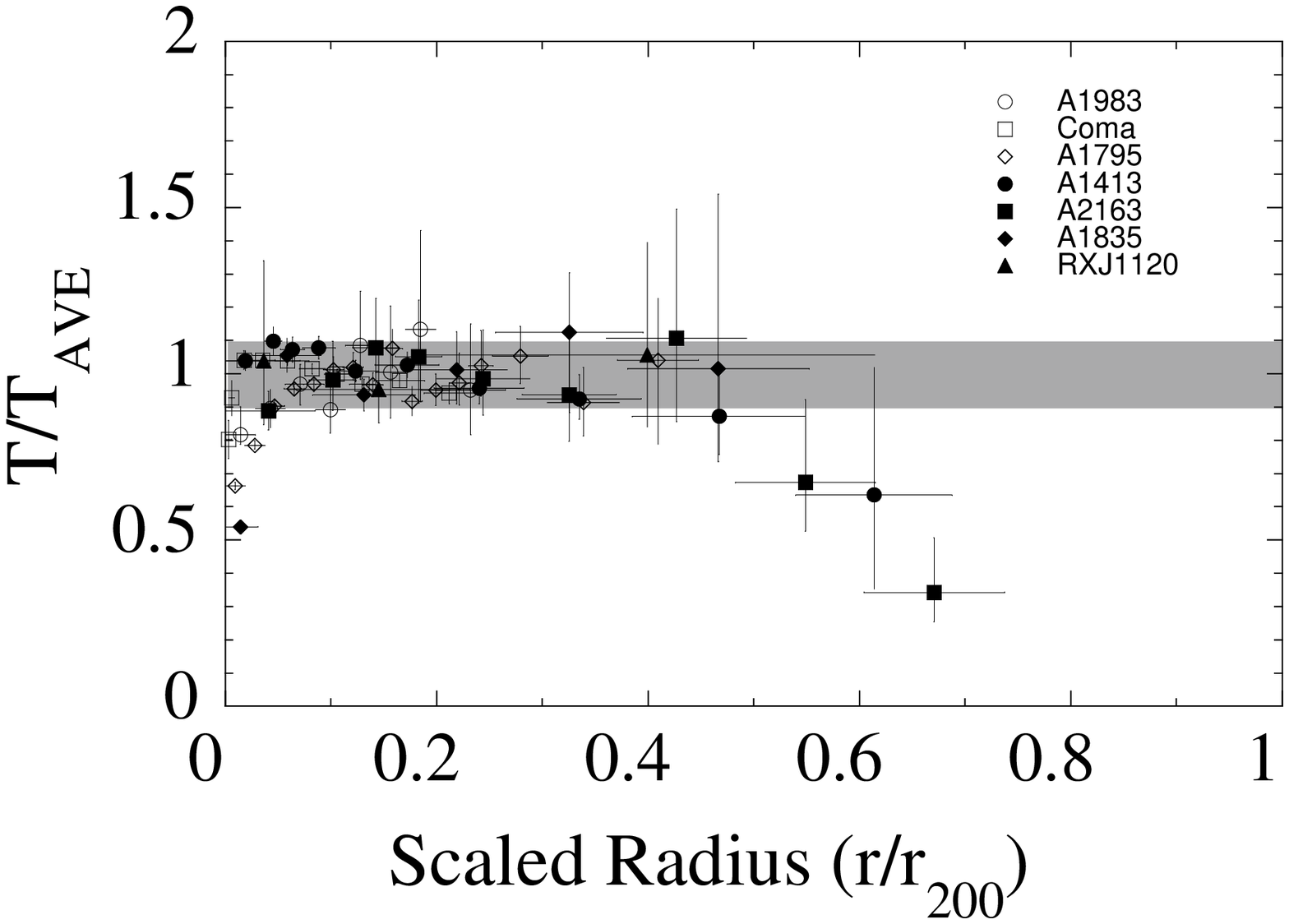}
    \caption{\footnotesize  Compilation of scaled $\kT$ profiles observed by \xmm. The temperature is scaled to the mean temperature outside the cooling flow region.  The data on Coma, A1795, A2163, RXJ1120, A1835, A1413,   and A1983 are from \citet{arn01a};  \citet{arn01b}; \citet{pratt01}; \citet{arn02}; \citet{maje02}, \citet{pratt02},  \citet{pratt03}, respectively.  The grey area corresponds to a variation of $\pm 10\%$ around unity.   }
              \label{fig:tprofs}%
    \end{figure}

In Fig.~\ref{fig:tprofs} we show the scaled temperature profiles of a haphazard sample of clusters which have been worked on by the Saclay group. Of these clusters, two display merger signatures (Coma, A2163), while the rest seem relatively relaxed. The average temperatures range between 2.1 and 14 keV, while the redshifts lie between $z=0.02$ and $z=0.6$. Despite the differing properties, the scaled temperature profiles are remarkably similar. What dispersion there is, is connected to the clear, well resolved drop in the central regions, the signature of a cooling core. The profiles are essentially isothermal (within $\sim  \pm 10\%$) up to $\sim 0.5 \rv$, after which there may be a decline, although we are presently limited by the small number of data points in this region. The great advantage of \xmm\ is, of course, that we actually have temperature measurements this far out from the cluster centre.

\subsection{Entropy profiles}

The entropy, defined, as is now customary, as $S = kT/n_e^{2/3}$,  is a fundamental characteristic of the ICM, because it is a probe of the thermo-dynamic history of the gas (e.g \citealt{voit03}). In the standard self-similar picture, the entropy, at any scaled radius, should scale simply as $S \propto h(z)^{4/3}T$. The pioneering works of \citet{ponm99} and  \citet{llyo00} have largely been superseded by the recent work of \citet{ponm03}, where they study the entropy profiles and scaling of 66 nearby systems observed by \asca\ and \rosat. These authors conclude that, while there is considerable dispersion particularly at the low mass end, the entropy measured at $0.1 \rv$ in fact scales as $S \propto T^{0.65}$, with cooler systems having systematically more entropy than expected.  This `entropy excess' has been the subject of a large amount of debate in the literature. Proposed explanations have included heating before or after collapse (from SNs or AGNs), radiative cooling  \ldots (\citealt{voit02}; \citealt{borg02} and references therein).

\begin{figure}[t]
     \hspace*{-1em} \includegraphics[width=7cm]{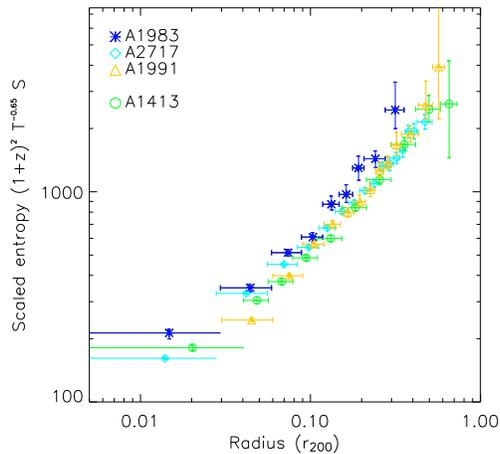}\vspace*{-1em}
   \caption{\footnotesize Scaled \xmm\ entropy profiles using $S \propto (1+z)^{-2} T^{0.65}$.
   }
              \label{fig:entropy}%
    \end{figure}

\begin{figure*}
   \hspace*{-1em}
  \includegraphics[width=14cm]{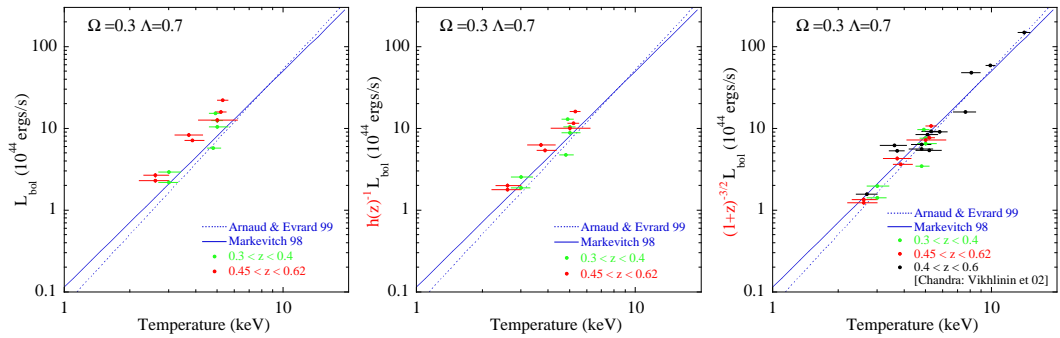}
    \caption{\footnotesize Correlation between bolometric luminosity and temperature ($\Omega_{\rm m}=0.3, \Lambda=0.7$). Blue lines: local relations from \citet{ae99} and \citet{mark98a}  for comparison.  Left panel: Data points for the medium $z$ (green points)  and high $z$ (red points) SHARC sample observed with \xmm. All the data points stand significantly above the local relation, indicating evolution. Middle panel: Luminosities scaled by $h(z)^{-1}$, removing the evolution expected in the standard self-similar model. There is a slight indication that the evolution is actually stronger . Right panel: Luminosities have been scaled by $(1+z)^{-3/2}$, according to the best fit evolution found by \citet{lumb03}. Black points: \chandra\ data from \citet{vikh02}. Note the excellent agreement between \chandra\ and \xmm.}
   \label{fig:LTz}
    \end{figure*}

In Fig.~\ref{fig:entropy} we show the \xmm\ entropy profiles of four clusters. Three (A1983, A1991 and A2717) are cool clusters ($\kT \sim 2.0-2.5$ keV; Pratt \& Arnaud, in prep.); the last is the massive cluster A1413 ($kT = 6.5$ keV; \citealt{pratt02}). The profiles have been scaled using the empirically determined $S \propto T^{0.65}$ relation. These are the first high quality profiles which range from $0.01$--$0.5 \rv$, and show the remarkable similarity of shape\footnote{We note that recent results from \chandra\ suggest far more dispersion in groups at about 1 keV \citep{sun03}.} and the excellent normalisation of the profiles with the adopted $S$---$T$ relation. The external slope of the entropy profiles is roughly consistent with the $S \propto r^{1.1}$ behaviour expected from shock heating.

These profiles allow us to rule out simple preheating models for the entropy excess. As shown in \citet{tozz01} and other works, preheating the gas leads inescapably to lower mass systems having large isentropic cores. It can be seen that the scaled core sizes of all of these systems is similar, and that there is no evidence for a large isentropic core in the cooler systems.

\section{Evolution of clusters}

The evolution of cluster properties contains key information to i) help disentangle the respective role of gravitational and various non-gravitational processes, and ii) to fully test the self-similar model of the DM collapse.

The standard self-similar model makes definitive predictions on the evolution of cluster properties. While they should keep the same internal structure as nearby clusters, more distant clusters should be denser, smaller and more luminous (see Sec. 1). ROSAT/ASCA observations gave the first indication that the self-similarity does hold at $z>0$.  The universal $EM$  profile appears to extend to $z\sim0.8$, with a redshift scaling consistent with the expectation for a $\Lambda$CDM cosmology  \citep{aan02}. These authors found significant evolution in the normalisation of the \LxT relation, consistent with the self-similar model (as have other studies considering this cosmology; \citealt{reic99}, \citealt{novi02}).  However these observations were highly biased towards massive systems, mostly clusters discovered by the EMSS, and their statistical quality was poor.  

Several X--ray samples of distant clusters have been
assembled using \rosat\ observations.  They are much larger and cover a
much wider mass range than the EMSS. With \xmm\ it
is now possible to make detailed studies of these clusters. 

We present here results from a follow-up of the clusters detected in the Bright and South SHARC survey (\citealt{rome00}; \citealt{burk03}).
Eight high z ($0.45<z<0.62$) clusters were observed in Guaranteed Time (\citealt{lumb03}, \citealt{arn02}) and  six  medium redshift ($0.3<z<0.4$) clusters in Open Time.  A combined analysis of the whole sample is in progress (Arnaud \etal, in prep.; see also \citealt{maje03}).  

These data definitively confirm a positive evolution in the normalisation of the \LxT relation (\citealt{lumb03}, Fig.~\ref{fig:LTz}). In the left panel of Fig.~\ref{fig:LTz},  we show the luminosities for the combined SHARC sample versus temperature. All the data points stand significantly above the local relation. \citet{lumb03} found that the normalisation of the \LxT relation for the high $z$ sample scales as $(1+z)^{1.54\pm0.26}$, consistent with the evolution found by \citet{vikh02} from \chandra\ data ($(1+z)^{1.5\pm0.3}$). The right panel illustrates this evolution: when each luminosity is scaled by $(1+z)^{-1.5}$,  the data points of the whole \xmm\ and \chandra\  samples become consistent with the local relation.  Note that this best fit evolution 
is stronger than the $h(z)$ evolution 
predicted by the standard self-similar model (see middle panel), although the effect is probably not very significant (work in progress). 
 
\begin{figure}[t]
 \hspace*{-1em}
  \includegraphics[width=7cm]{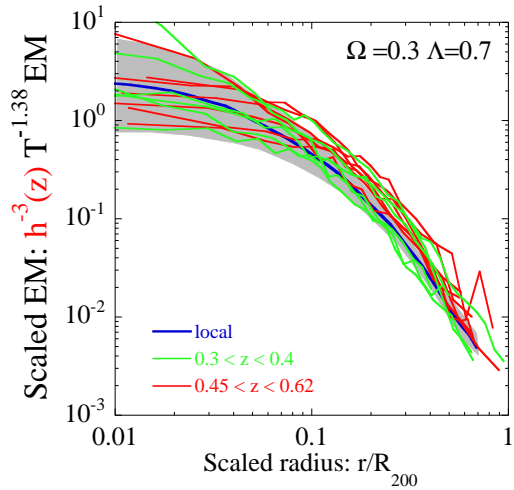}
    \caption{\footnotesize The scaled $EM$ profiles of medium and high redshift SHARC clusters (green and red lines) observed with  \xmm, compared to  the mean scaled profile of nearby hot clusters (full line, the grey area 
is the scatter around the mean, see Fig.~\ref{fig:emprof}). }
   \label{fig:EMscz} 
    \end{figure}

We show in Fig.~\ref{fig:EMscz} the corresponding $EM$ profiles,  scaled according to the standard evolution:  	$EM_{\rm sc} =  h(z)^{-3} T^{-1.38} EM$ (the empirical slope of the \EMT relation derived by \citet{neum01}, is assumed not to evolve). 
The profiles are traced up to similar, or even larger, radii than the mean profile of nearby clusters observed with \rosat. There is good agreement between individual profiles and the local mean profile (taking into account the local dispersion and errors); the self-similarity of form up to high $z$ is confirmed. However there is a hint of a systematic discrepancy: most high $z$ profiles lie above the above local curve (the effect is stronger for the highest $z$ sample). This is again an indication for stronger evolution, as found from the \LxT relation. 
%

\section{Conclusions}
The picture that emerges from the \xmm\ observations presented here is that clusters do form a self-similar population.  

The self-similarity of shape for the gas component (density and $T$ profiles beyond the cooling core, entropy profiles) extends down to surprisingly low temperature and up to high redshift. We begin to have evidence that the dark matter  profile is universal and presents a central cusp, as predicted by numerical simulations.  Scaling laws do relate various physical properties with $z$ and $T$, with a slope of the $\MT$ relation as expected. This supports the validity of the current approach for the modeling of the {\it dark matter} collapse.  

However, the {\it gas}  scaling properties are significantly different from expected in the simple model. The steepeer  $\EMT$ and shallower $\ST$ relations for nearby clusters is confirmed at all radii and down to low temperature, and are consistent with the steepening of the $\MgT$ and $\LxT$ relations.  As all these quantities are related to the overall gas content, $\overline\rhog$, this suggests that the fundamental deviation from the simplest model is an increase of  $\overline\rhog$ with $T$ (or mass), or, equivalently, an increase of the gas mass fraction with $T$ (if $M \propto T^{1.5}$). There is also some indication of stronger evolution than expected. Furthermore, current adiabatic numerical simulations
do not predict the correct normalisation of the local \MT relation, suggesting that the gas profile shape is not correctly reproduced.

These results confirm  that the physics governing the baryonic component is still far from understood.  Recent modeling, including gas cooling and/or non gravitational heating, has shown that these  processes can affect the scaling relations. However, the self-similarity of shape is a strong new constraint, and pure pre-heating models can already be ruled out. 

These results must still be taken with caution. In spite of the quality of the data, they are based on a small data set of clusters, and should be considered as 'test cases'.  Observations of large samples, with unbiased mass and redshift sampling, is essential i) to confirm the self-similarity of form, ii) to firmly establish the slope and normalisation of the $\MT$ relation and the intrinsic scatter in cluster properties, and iii) fully to assess cluster evolution, including possible evolution of the scaling laws slopes. 

\begin{acknowledgements}
MA thanks D. Lumb and the XMM-$\Omega$ project collaboration for providing the $S_{\rm X}$ profiles published in \citet{lumb03}.  
\end{acknowledgements}
\vspace{-0.5cm}
\bibliographystyle{aa} 

\begin{thebibliography}{} 

\bibitem[Allen \etal (2002)]{allen02} 
	Allen S.W., Schmidt, R.W., Fabian, A.C., 2002, \mnras, 335, 256

\bibitem[Arnaud \& Evrard (1999)]{ae99} 
	Arnaud M. ,  Evrard A.E., 1999, \mnras, 305, 631 

\bibitem[Arnaud \etal (2001a)]{arn01a}  
	Arnaud, M., Aghanim, N., Gastaud, R. \etal 2001a, \aap, 365, L67  \bibitem[Arnaud \etal (2001b)]{arn01b}  
	Arnaud, M., Neumann, D., Aghanim, N., \etal 2001b, \aap, 365, L80 

\bibitem[Arnaud (2002)]{arn02r}
 	Arnaud M., 2002, ASPConfSer, 268 157

\bibitem[Arnaud \etal (2002a)]{aan02}  
	Arnaud  M., Aghanim N., Neumann D., 2002a, \aap, 389, 1 

\bibitem[Arnaud  \etal (2002b)]{arn02} 
	Arnaud M., Majerowicz S., Lumb, D. \etal, 2002b, \aap, 390, 27

\bibitem[Bertschinger (1985)]{bert85} 
	Bertschinger E. 1985, \apjs, 58,39

\bibitem[Borgani \etal (2002)]{borg02}
	 Borgani, S., Governato F., Wadsley J. \etal 2002, \mnras, 336, 409 

\bibitem[Bryan \&Norman (1998)]{bryan98} 
	Bryan, G.L., Norman, M.L. 1998, ApJ, 495, 80 

\bibitem[Burke \etal (2003)]{burk03}  
	Burke, D.J., Collins C.A., Sharples, R.M. \etal 2003, \mnras, 341, 1093

\bibitem[Castillo-Morales \& Schindler (2003)]{cast03}
	Castillo-Morales A., Schindler S., 2003, \aa, 403, 433

\bibitem[Cavaliere \etal(1999)]{cav99}  
	Cavaliere A., Menci N., Tozzi P., 1999, \mnras, 308, 599 

\bibitem[De Grandi \& Molendi (2002)]{dm02} 
	De Grandi, S., Molendi, S. 2002, \apj, 567, 163

\bibitem[{Ettori \etal  (2002)}]{etto02}
	Ettori, S., De Grandi, S., Molendi, S. 2002, \aa, 391, 841 

\bibitem[{Evrard \& Henry  (1991)}]{eh91} 
	Evrard A. \& Henry J.P., 1991, \apj, 383, 95

\bibitem[{Evrard \etal (1996)}]{emn96} 
	Evrard A.~E., Metzler C.~A.,  Navarro, J.~F., 1996, \apj, 469, 494 

\bibitem[{Evrard \& Gioia  (2002)}]{eg02}  
	Evrard A., \& Gioia I.  2002 Ast\& Sp.  Sc.  Lib.  272, 253

\bibitem[{Finoguenov \etal (2001)}]{fino01}  
	Finoguenov A., Reiprich T. H., B\"{o}hringer H., 2001, \aap, 368, 749 

\bibitem[Horner \etal (1999)]{horn99}  
	Horner D., Mushotzky R., Scharf C., 1999, \apj, 520, 78  

\bibitem[Irwin \& Bregman (2000)]{ib00}  
	Irwin J., Bregman J.,  2000, \apj, 538, 543

\bibitem[Lloyd-Davies \etal (2000)]{llyo00} 
	 Lloyd-Davies, E.J., Ponman, T.J., Cannon, D.B. 2000, MNRAS, 315, 689 

\bibitem[Lumb \etal (2003)]{lumb03}
	Lumb D.H., Bartlett J., Romer A.K. \etal, 2003, \aa, submitted, astro-ph/0311344
	
\bibitem[Markevitch (1998)]{mark98a} 
	Markevitch, M. 1998, \apj,  504, 27 

\bibitem[Markevitch \etal (1998)]{mark98b} 
	Markevitch, M., Forman W., Sarazin C., Vikhlinin  A. 1998, \apj, 503, 77 

\bibitem[Majerowicz \etal(2002)]{maje02} 
	Majerowicz, S., Neumann, D.M., Reiprich, T.H. 2002, \aap, 394, 77 

\bibitem[Majerowicz (2003)]{maje03} 
	Majerowicz, S., 2003, phD Thesis 

\bibitem[Mohr \etal  (1999)]{mohr99} 
	 Mohr J.J., Mathiesen B., Evrard A.E., 1999, \apj, 517, 627 

\bibitem[Moore \etal (1999)]{moor99} Moore, B., Quinn, T., Governato, F., Stadel, J., \& Lake, G.\ 1999, \mnras, 310, 1147 

\bibitem[Navarro \etal (1995)]{nfw95} 
	Navarro J.F., Frenk C.S., White S.D.M., 1995, \mnras, 274, 720 

\bibitem[Navarro \etal (1997)]{nfw97} 
	Navarro J.F., Frenk C.S., White S.D.M., 1997, \apj, 490, 493

\bibitem[Neuman \& Arnaud (1999)]{neum99} 
	Neuman D.M., Arnaud, M. 1999, \aap, 348, 711  

\bibitem[Neuman \& Arnaud (2001)]{neum01} 
	Neumann, D.M., Arnaud, M. 2001, \aap, 373, L33

\bibitem[Nevalainen \etal (2000)]{neva00}
	Nevalainen, J., Markevitch, M., Forman, W. 2000, \apj, 532, 694

\bibitem[Novicki \etal (2002)]{novi02}
	Novicki, M.C., Sornig M., Henry J.P., 2002, \aj, 124, 2413    

\bibitem[{Ponman \etal (1999)}]{ponm99}  
	Ponman T., Cannon D., Navarro, J. 1999, Nature, 397, 135

\bibitem[{Ponman \etal (2003)}]{ponm03} 
	Ponman, T.J., Sanderson, A.J.R., Finoguenov, A., 2003, MNRAS, 343, 331

\bibitem[{Pratt \etal (2001)}]{pratt01}  
	Pratt, G.W., Arnaud, M., Aghanim, N., 2001, Proc.  XXXVI Rencontres de Moriond: Galaxy Clusters and the High-Redshift Universe, eds.  D.M. Neumann, F. Durret and J. Tr\^{a}n Thanh Van: astro-ph/0105431. 

\bibitem[{Pratt \& Arnaud (2002)}]{pratt02} 
	Pratt G.~W., Arnaud M.,  2002, \aap, 394, 375

\bibitem[{Pratt \& Arnaud (2003)}]{pratt03} 
	Pratt, G.~W.,  Arnaud M.,  2003, \aap, 408, 1

\bibitem[{Pointecouteau \etal (2003)}]{pointeco03} 
	Pointecouteau, E.~\& Arnaud, M., Kaastra, J.~, dePlaa, J. 2003, \aa, submitted

\bibitem[Reichart \etal (1999)]{reic99} 
	Reichart, D.E., Castander, F.J., Nichol, R.C.  1999, ApJ, 516, 1 

\bibitem[Romer \etal (2000)]{rome00} 
	Romer, A. K., Nichol, R. C., Holden, B. P. \etal, 2000, \apjs,  126, 209 

\bibitem[{Sanderson \etal (2003)}]{sand03}
	Sanderson A.J.R., Ponman T., 	Finoguenov A., \etal , 2003, \mnras, 340, 989

\bibitem[Sun \etal (2003)]{sun03} 
	Sun M., Forman W., Vikhlinin A., \etal  2003, \apj, 598, 250

\bibitem[Tozzi \& Norman (2001)]{tozz01} 	Tozzi, P., \&  Norman, C., 2001, ApJ, 546, 63 	
\bibitem[{Vikhlinin \etal (1999)}]{vikh99} 
	Vikhlinin A., Forman, W., Jones, C. 1999, \apj, 525, 47 

\bibitem[{Vikhlinin \etal (2002)}]{vikh02} 
	Vikhlinin A., VanSpeybroeck, L., Markevitch M., \etal  2002, \apj, 578, L107 

\bibitem[Voit \etal (2002)]{voit02}
	 Voit, G.M., Bryan G.L., Balogh, M.L., Bower, R.G. 2002, \apj, 576, 601

\bibitem[Voit \etal (2003)]{voit03}
	 Voit, G.M., Balogh, M.L.,   Bower R.G., \etal  2002, \apj, 593, 272	 
\bibitem[{White (2000)}]{whit00}  
	White D.A. 2000, \mnras, 312, 663

\bibitem[Xu \etal (2001)]{xu01}
	Xu H., Jin G., Wu X.P. 2001, \apj, 553, 78

\end{thebibliography}

\end{document}